# Characterizing Niobium Nitride Superconducting Microwave Coplanar Waveguide Resonator Array for Circuit Quantum Electrodynamics in Extreme Conditions

Paniz Foshat[1], Paul Baity[1], Sergey Danilin[1], Valentino Seferai[1], Shima Poorgholam-Khanjari[1], Hua Feng[1], Oleg A. Mukhanov[2], Matthew Hutchings[2], Robert H. Hadfield[1], Muhammad Imran[1], Martin Weides[1], and Kaveh Delfanazari[1,*]

[1] *Electronics and Nanoscale Engineering Division, James Watt School of Engineering, University of Glasgow, Glasgow, UK*
[2] *SeeQC, UK*

\* kaveh.delfanazari@glasgow.ac.uk  Dated: 01062023

*Abstract:*

The high critical magnetic field and relatively high critical temperature of niobium nitride (NbN) make it a promising material candidate for applications in superconducting quantum technology. However, NbN-based devices and circuits are sensitive to decoherence sources such as two-level system (TLS) defects. Here, we numerically and experimentally investigate NbN superconducting microwave coplanar waveguide resonator arrays, with a 100 nm thickness, capacitively coupled to a common coplanar waveguide on a silicon chip. We observe that the resonators' internal quality factor ($Q_i$) decreases from $Q_i \sim 1.07 \times 10^6$ in a high power regime ($<n_{ph}> = 27000$) to $Q_i \sim 1.36 \times 10^5$ in single photon regime at temperature $T = 100$ mK. Data from this study is consistent with the TLS theory, which describes the TLS interactions in resonator substrates and interfaces. Moreover, we study the temperature dependence internal quality factor and frequency tuning of the coplanar waveguide resonators to characterise the quasiparticle density of NbN. We observe that the increase in kinetic inductance at higher temperatures is the main reason for the frequency shift. Finally, we measure the resonators' resonance frequency and internal quality factor at single photon regime in response to in-plane magnetic fields $B_{//}$. We verify that $Q_i$ stays well above $10^4$ up to $B_{//} = 240$ mT in the photon number $<n_{ph}> = 1.8$ at $T = 100$ mK. Our results may pave the way for realising robust microwave superconducting circuits for circuit quantum electrodynamics (cQED) at high magnetic fields necessary for fault-tolerant quantum computing, and ultrasensitive quantum sensing.

# 1. Introduction

Superconducting coplanar waveguide resonators, or CPWs, play an essential role in circuit quantum electrodynamics (cQED) [1-3], superconducting quantum sensing [4-6], quantum metrology [7, 8], and quantum computing [9, 10]. The electric field in CPWs can be confined inside the resonator, making them an appropriate choice for coupling between quantum devices. A further benefit of CPWs is their simple geometry, which makes them a fundamental component of quantum circuits. In this regard, it is crucial to maintain a high internal quality factor ($Q_i$) of CPWs in quantum circuits to minimise unwanted losses that affect qubit gate fidelities and qubit relaxation times [11, 12] to achieve robust and fault-tolerant quantum computers [10, 13].

Recently, attention has been on identifying the loss mechanisms contributing to the quality factor of the CPWs [14]. Recent results show that two-level systems (TLSs) and quasiparticle losses are mostly responsible for quantum circuits' decoherence [15-18]. TLS loss appears when the defects in amorphous material (such as the substrate of quantum circuits and oxides growth on the superconductor) are coupled with the electric field of CPWs. These TLS defects are mostly confined to the ground states at millikelvin temperatures and low powers. They absorb energy from CPWs and dissipate heat to the resonator when defects jump to excited energy states [19]. Accordingly, circuit engineering techniques have been developed on topics of finding low-loss substrate and superconductor materials [20, 21], optimising circuits geometry [22-24] and improving micro and nanofabrication techniques [25, 26]. Several research groups have pursued these methods to improve the quality factor in the CPWs and achieved a high internal quality factor's CPWs in the quantum devices [27-31].

Moreover, there has been considerable interest in the characterisation of the quasiparticle loss [32-35], another cause of loss in CPW. An active area of research is to find the dynamic of the quasiparticle effect on superconducting quantum devices [36-38], which can host Andreev bound states (ABSs) [39-41] and Majorana bound states (MBSs) [39, 42]. Emerging MBSs in hybrid superconductor-semiconductor devices require an application of magnetic fields while maintaining the superconductivity [43]. Therefore, topological superconductivity can potentially appear in a hybrid circuit with superconductors, such as niobium titanium nitride (NbTiN) [44, 45], tantalum (Ta) [46], lead (Pb) [47], and niobium nitride (NbN) with their relatively high critical magnetic fields, and also for this purpose, it will be important to characterise the microwave superconducting circuits losses in the presence of the magnetic field.

A main challenge in superconducting circuits operating in extreme conditions, e.g., in the presence of magnetic fields, is achieving high $Q_i$ resonance since changes in the magnetic flux density in superconductors cause suppression of superconductivity or a phase transition from superconductor states to normal states. Moreover, in type II superconductors, the superconductor enters to the mixed states when the magnetic field is applied, leading to vortices pinning and flux creep. The following issues need to be addressed to preserve high $Q_i$ resonators based on the type II superconductors [30]. First, the $Q_i$ becomes more vulnerable to even small magnetic fields due to the creation of Abrikosov vortices: supercurrent regions that circulate around a non-superconducting core [48]. Second, type II superconductors have a lower coherence length in comparison with type I superconductors; therefore, energy losses have a higher value at single photon regimes. The possible solutions to overcome these challenges are to reduce the geometry of the resonator to below magnetic field penetration depth or to define lithographically artificial defect sites to pin the vortices [48-51].

In this work, we first report the design and fabrication of NbN CPWs arrays on two silicon chips whose frequency resonance window is between 4 GHz and 8 GHz. Following this, we discuss characterisation and analyses of the frequency shift $\Delta f$ and internal quality factor $Q_i$ of the CPWs in response to sweeping temperature, microwave power and magnetic field. We further verify that the $Q_i$ of CPW in one of the chips increases from $Q_i \sim 1.36 \times 10^5$ in single photon regime to $Q_i \sim 1.07 \times 10^6$ at photon number $<n_{ph}> = 2.7 \times 10^4$, at $T = 100$ mK. Indeed, we find that the $\Delta f$ of the resonator changes when the magnetic field is applied at different temperatures. Moreover, we show that $Q_i$ stays above $Q_i \sim 10^4$ at $<n_{ph}> = 1.9 \times 10^5$, $<n_{ph}> = 6$ and $<n_{ph}> = 1.8$ when parallel magnetic fields $B_\parallel < 240$ mT are applied to the chips at $T = 100$ mK. These results show that the NbN resonators are promising superconducting material candidates for hybrid quantum electrodynamic circuits requiring magnetic fields.

**2. Method**

2.1 Design and Fabrication

Our superconducting microwave circuits consist of an array of microwave resonators capacitively coupled to a common coplanar waveguide and fabricated on a 100 nm thin NbN film sputtered on a silicon chip. The silicon substrate is dipped in Buffered Oxide Etchant (BOE) to strip the native $SiO_x$ from the substrate's surface to achieve a clean substrate-metal interface. Immediately after removing the $SiO_x$ layer, the chip was placed in the load lock of an MP 600 S Plassys sputter system to prepare for NbN metal deposition [52]. After the load lock pumps down and reaches a low pressure, the sample was transferred to the main chamber to reactively sputter 100 nm of NbN. In the main chamber, by sputtering niobium (Nb) for 15 minutes with 25 sccm of Argon (Ar) and 3.0 sccm of Nitrogen at the base current of 0.85 A, 100 nm NbN film was formed on top of the 525 µm silicon substrate. After removing samples from the load lock of the sputtering machine, ZEP resist was spun and baked at 180 °C on hotplates for four minutes. The resonators were patterned with standard e-beam lithography followed by $CF_4$/Ar anisotropic dry etch. The resonator length was chosen to have a fundamental frequency mode between 4 – 8 GHz.

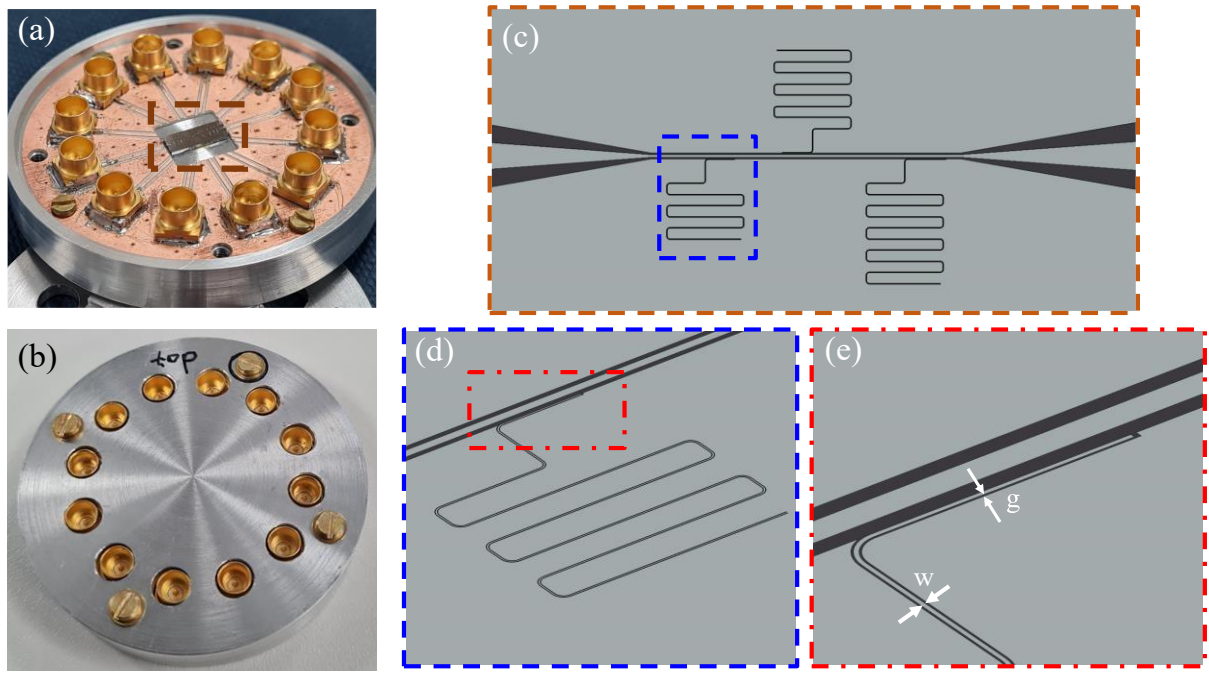

**Figure 1.** NbN based superconducting microwave coplanar waveguide resonators. (a) Optical image of sample 1 which is wire-bonded in the sample holder. (b) Packaged chip for cryogenic microwave spectroscopy. (c) A top-view schematic of the superconducting circuit used in the experiments shows three capacitively coupled resonators multiplexed to a common coplanar waveguide (brown dashed line area in Fig. (a)). The length of resonators from left to right are $L_1$ = 4.688 mm, $L_2$ = 6.250 mm, $L_3$ = 7.900 mm. Here, the thickness of NbN film, and silicon substrate is 100 nm, and 525 µm, respectively. (d) Magnified schematic of resonator 1 (blue dashed line in Fig. 1(c)). (e) The magnified area shows the resonator's gap and width position (red dashed line in Fig. 1(d)).

Two different types of samples are fabricated and measured, each containing three resonator arrays with 4 µm width, and 2 µm gap, multiplexed to a single feedline. Figures 1(a), and (b) show a wire-bonded superconducting CPW chip, and the packaged chip, respectively. A top-view schematic of one resonator is shown in Fig.1(c). Moreover, a zoom-in view of the resonator (R1) and the magnified area showing the gap and width position are schematically depicted in Fig.1(d) and Fig.1(e), respectively. More details of the fabricated superconducting CPW circuits in this study are provided in Table 1.

Table 1. NbN superconducting microwave coplanar waveguide resonator array design detail.

| Sample | Resonator Width | Resonator Gap | R1 Length | R2 Length | R3 Length |
|---|---|---|---|---|---|
| 1 | 4 µm | 2 µm | 4.688 mm | 6.250 mm | 7.9 mm |
| 2 | 4 µm | 2 µm | 4.5 mm | 5.5 mm | 6.5 mm |

2.2 Experimental setup

The sample box (Fig.1(b)) was mounted in two different cryogenic setups to collect the data presented in this paper: 1) an adiabatic demagnetization refrigerator (ADR) with a base temperature of $T = 30$ mK (see the fridge setup in FIG.S1. (a)). 2) a dilution refrigerator (DR) with a base temperature of $T = 20$ mK (see FIG.S1. (b)). The complex microwave transmission spectroscopy ($S_{21}$) was performed with a MS4642B Keysight Vector Network Analyzer (VNA) by sending signals to ports on the NbN transmission line. In the ADR setup, the input line of the VNA is heavily attenuated (-40 dB) at room temperature. Then, VNA signals were attenuated by -20 dB at $T = 70$K, 4K, and 0.5 K stages before reaching the transmission line of the sample, suppressing thermal noise and allowing the resonator to reach into the single photon regime. At the $T = 4$K stage, after the signal passes through the sample box, an isolator is placed to prevent contamination of transmission signals from spurious reflections back to the transmission line. Then, the signal was amplified by a high electron mobility transistor (HEMT) which provided +40 dB amplification. Moreover, an additional +45 dB amplification is provided with a room-temperature amplifier. RF signals generated by the VNA propagate through the ADR before reaching the CPW transmission line, where they excite the resonators. The signal power levels range from -30dBm to 15dBm (see FIG.S1. (a)).

We note that, for magnetic field measurements, we mounted our sample inside a Solenoid (NbTi superconducting magnets) in the ADR such that the NbN transmission line is parallel to the applied magnetic fields. The current in the Solenoid sweeps between -10 A to 10 A, leading to a magnetic field sweep between -800 mT to 800 mT [51]. In this work, we present the microwave power and temperature sweep of sample 2 which are measured in a dilution refrigerator (DR) and the ADR refrigerator, and the magnetic field sweep of sample 1 measured in the ADR refrigerator. The DR cryogenic setup detail is shown in FIG.S1. (b).

2.3 Numerical methodology

The extraction of resonator parameters such as resonance frequency ($f_r$), internal quality factor ($Q_i$), coupling quality factor ($Q_c$), and loaded quality factor ($Q_l$) from $S_{21}$ measurements, is a crucial tasks in the characterisation and design of resonate circuit. To pursue this goal different types of RLC circuit models have been used in previous research based on amplitude or phase of $S_{21}$ [28, 53-56]. Modern methods typically utilize the full complex scattering data provided by VNA as opposed to focusing on the power or amplitude to extract resonator parameters [28, 57, 58]. Using entire complex data results in a better signal-to-noise ratio (SNR) and a more accurate determination of resonator parameters. This article uses a notch-type resonator model to extract resonance frequency and quality factors of the NbN microwave superconducting resonators. The $S_{21}$ of notch-type resonators can be evaluated with the following equation [58]:

$$S_{21}^{notch}(f) = \underbrace{a\, e^{i\alpha} e^{-2\pi i f \tau}}_{\text{Environmental effect}} \times \underbrace{\left[1 - \frac{\left(\frac{Q_l}{|Q_c|}\right) e^{i\varphi}}{1 + 2iQ_l\left(\frac{f}{f_r} - 1\right)}\right]}_{\text{Ideal resonator}} \quad (1)$$

The second part of Eq. (1) characterises an ideal notch-type resonator where $f$ indicates the probe frequency, and $\varphi$ quantifies the impedance mismatch. In more detail, $\varphi$ indicates the difference between input and output impedance at the ports of the resonator. In an ideal notch-type RLC circuit, the circle (Im($S_{21}$) and Re($S_{21}$) circle plot) intersects the real axis at unity, which corresponds to a probe frequency at $f \to \pm\infty$ called off-resonant point.

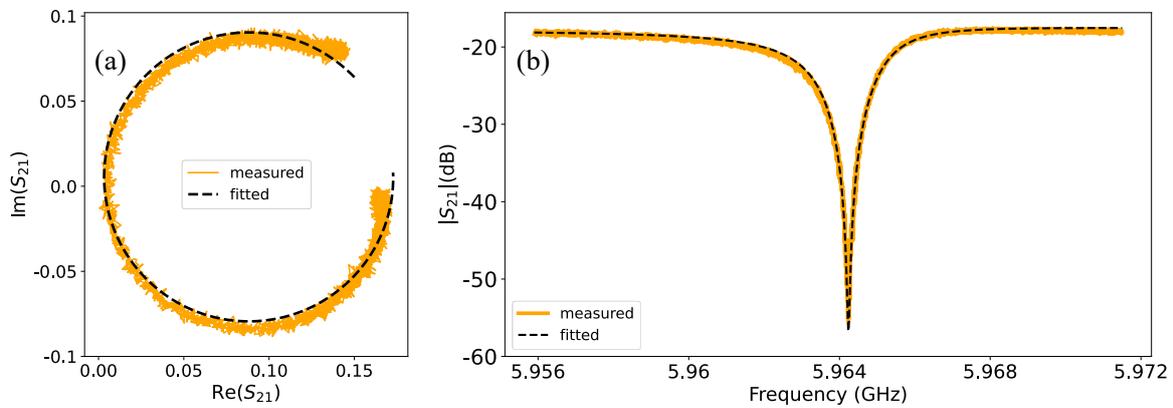

**Figure 2.** (a) Parametric plot and fitted Im ($S_{21}$) and Re ($S_{21}$), (b) Measured and calculated frequency spectrum of sample 2 at fundamental frequency $f_r$ = 5.9643 GHz at $T$ = 26 mK, same data in (a).

Note that the diameter ($d$) of the circle comes from $d = \frac{Q_l}{|Q_c|}$. However, in experimental measurements from VNA, a rotation in the phase and a change in the amplitude of the circle have been observed. To overcome the above issue, the first term in Eq. (1) has been added to describe the environmental noise. First, an amplitude $a$ shows the cable damping effect in $S_{21}$, a phase shift $\alpha$ shows an extra rotation in the phase of $S_{21}$ coming from cable length, and finally, an electronic delay $\tau$ which changes the initial phase in signal input $S_{21}$ to a different value from zero. The length of the cable and the finite speed of light cause an electronic delay. Besides, with a circuit model we also can drive the photon number inside the resonator with the below formula [59]:

$$P_{in} = P_{trans} + P_{reflection} + P_{loss} \qquad (2)$$

$$P_{loss} = P_{in}(1 - |S_{21}|^2 - |S_{11}|^2) \qquad (3)$$

$$<n_{ph}> = Q_i \times \frac{P_{loss}}{\hbar\omega_0^2} = Q_i \times \frac{P_{in}}{\hbar\omega_0^2}\left(2 \times Q_L \times \frac{Q_c - Q_L}{Q_c^2}\right) \qquad (4)$$

Where $n_{ph}$ is the photon number, $P_{in}$ is the input power, $P_{trans}$ is the transmitted power through the feedline, $P_{loss}$ is the source of losses, $\hbar$ is plank constant and $\omega_0$ is resonance angular frequency of resonator. A typical result of such a fit is shown in Fig. 2, where we show the measured and calculated frequency spectrum of sample 2 at fundamental frequency $f_r$ = 5.965 GHz at $T$ = 26 mK. Indeed, we investigate $Q_i$ relation with power in the DR cryogenic setup to confirm the reproducibility of measurements taken in the ADR and cover a higher range of photon numbers.

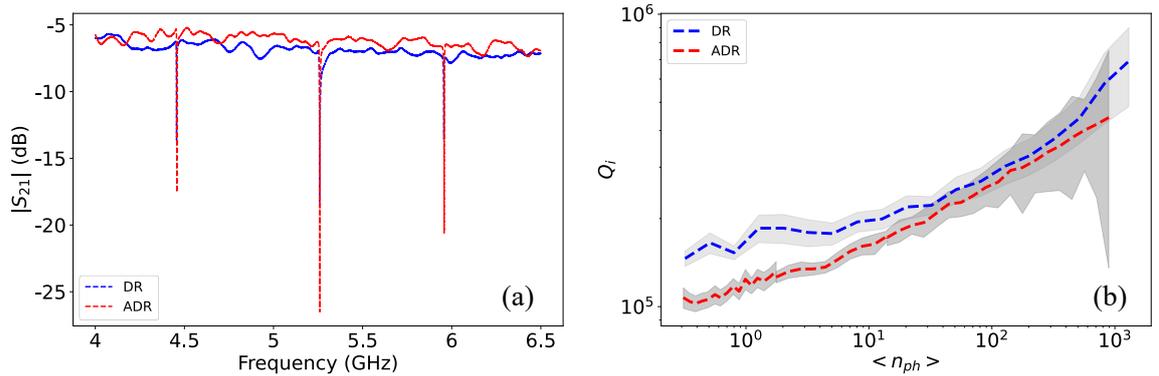

**Figure 3.** (a) Frequency spectrum of sample 2 measured in two different cryogenic setups: the ADR (red dashed-line), and DR (blue dashed-line). A good agreement between the two measurements can be seen at $T$ = 200 mK. (b) Measured $Q_i$ at fundamental frequency $f_r$ =5.96 GHz in the power range of $P_{in}$ = -150 dBm to -100 dBm at $T$ = 200 mK in two different setups.

Figure 3(a) shows the frequency spectrum comparison, confirming a good agreement between measurements taken in two different DR and ADR cryogenic setups. In Fig 3. (b), we compare extracted $Q_i$ and $< n_{ph} >$ taken in the ADR and DR cryogenic setup at power range -100 dBm $< P_{in} <$ -150 dBm at $T$ = 200 mK. The results confirm that both measurement setups are in accordance with each other during power sweep measurements. In the following chapter, we characterise kinetic inductance, TLS loss, and quasiparticle loss.

## 3. Results

## 3.1 Kinetic inductance characterisation

In the simplest case of capacitively coupled superconducting quantum circuits, modelling the resonator as an $LC$ circuit, the coupling strength $g$ between the feedline and resonator will be proportional to the impedance of the quantum circuits. In the quarter wavelength resonators impedance $Z$ and resonance frequency $f_r$ of the $LC$ circuit can be extracted by the equation $Z = \sqrt{\frac{(L_l+L_k)}{C_l}}$ and $v_{ph} = \frac{2\pi f_r}{k_n}$ where $k_n = \frac{\pi n}{2l}$ and $v_{ph} = \frac{1}{\sqrt{C_l(L_l+L_k)}}$. Here, $L_l$, $L_k$ and $C_l$ are geometrical inductance, kinetic inductance, and geometrical capacitance per unit length of transmission line, respectively. Moreover, $n$ emphasizes on the resonance frequency number ($n = 1$ for the fundamental mode) [28].

Therefore, characterising the kinetic inductance of superconducting circuits to identify coupling strength and resonance frequency is an essential step toward robust quantum technologies. Moreover, the resonators' kinetic inductance helps us to evaluate quasiparticle loss and frequency shift in superconducting circuits. More details are discussed in section 3.3. For obtaining kinetic inductance, we first use conformal mapping techniques [60] to extract the geometrical inductance $L_l = 4.1367 \times 10^{-7}$ H/m and the capacitance $C_l = 1.6803 \times 10^{-10}$ F/m. We obtain kinetic inductance $L_k = 4.464 \times 10^{-8}$ H/m, which shows lower kinetic inductance compared to previous reports on NbN coplanar waveguide resonators and thin NbN sheets [28, 61, 62]. The following section investigates the effects of power, temperature, and magnetic field on NbN's CPW.

## 3.2 Power dependence $Q_i$

Knowing how sweeping power and temperature affect material defects' behaviour can help us make more robust quantum circuits. To achieve this, we study the resonator's $Q_i$ to an average photon number inside the resonator at three different temperatures: at (i) $T = 26$ mK, (ii) $T =$

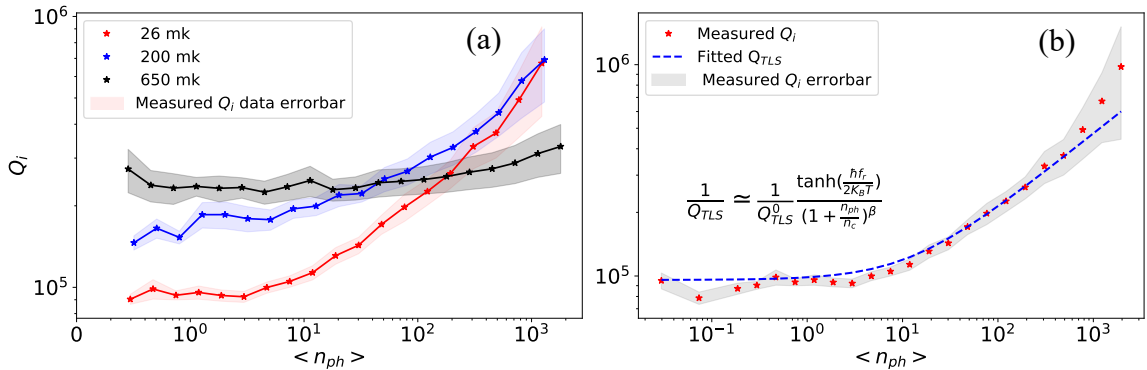

**Figure 4.** (a) Measured $Q_i$ at fundamental frequency $f_r = 5.96$ GHz in the power range $P = -150$ dBm to $-100$ dBm at $T = 26$ mK, 200 mK and 650 mK. The shaded area corresponding to deviation/error from the actual data. (b) $Q_i$ versus photon number at $T = 26$ mK fitted with the theoretical model of TLS. These data are measured for sample 2 in the DR.

200 mK, and (iii) $T$ = 650 mK. This is done in order to identify the sources of energy loss of the resonator at different temperatures. We split measured losses into three terms, as shown in equation (5). Here, $\delta_{TLS}$ describes TLS loss, $\delta_{qp}$ is quasiparticle loss, $\delta_B$ is vortex and magnetic field loss, and $\delta_0$ is power and temperature independent loss [33]

$$\delta = \delta_{TLS}(T,P) + \delta_{qp}(T) + \delta_B(B) + \delta_0 \tag{5}$$

TLS defects in the ground state are the dominant cause of increasing loss at sub-millikelvin temperatures since TLS defect's jumping to the excited states induces heating dissipation in the resonator [19]. Meanwhile, the superconducting energy gap is larger than the thermal noise of the system at temperatures $\Delta \gg k_B T$, where $\Delta$ is superconducting gap, $k_B$ is boltzman constant, so Cooper pair breaking is less likely to happen. Therefore, quasiparticle density becomes negligible. In this case, the loss rates can be characterised [21]:

$$\delta_{TLS}(T,P) = \frac{1}{Q_{TLS}} = \frac{1}{Q_{TLS}^0}\frac{\tanh\left(\frac{hf_r}{2K_BT}\right)}{(1+\frac{n_{ph}}{n_c})^\beta} \tag{6}$$

where $f_r$ is resonance frequency, $n_{ph}$ is the average photon number, $n_c$ is critical photon number, and $\frac{1}{Q_{TLS}^0} = \delta_{TLS}^0$ is TLS loss at zero power and temperature.

As shown in Fig. 4, in $T$ = 26 mK, $Q_i$ saturates at $10^5$ in single photon regime, and as power rises it goes to surpasses $10^6$. We find an agreement between TLS theory and measurement results, as shown in Fig. 4(b). We find that $Q_{TLS}^0$, $n_c$ and $\beta$ are 9.5102× $10^4$, 13 and 0.35, respectively. In comparison with previous measurements [27-30, 63] we achieve a $Q_i$ of around one million at only < $n_{ph}$ > = 2000, at $T$ = 26 mK, which suggests a good improvement in the development of high-$Q$ superconducting CPWs based on thin NbN films.

In comparison, at higher temperatures, TLS defects saturate in a mixture containing equal proportions of ground and excited states; therefore, an energy exchange happens between energy states, reducing the energy loss from TLS defects. However, the energy gap is near the required Cooper pair-breaking energy in this temperature range. Decreasing the superconducting energy gap leads to an increase in quasiparticle density in the resonator, which is the primary cause of energy loss at this temperature range. The quasiparticle loss can be characterised by the below model [64, 65]:

$$\frac{1}{Q_i} \sim \frac{1}{Q_{qp}} = \frac{\alpha}{\pi}\sqrt{\frac{2\Delta}{\hbar\omega_r}}\frac{n_{qp}(T)}{D(E_F)\Delta} \tag{7}$$

Where $\alpha$ is the ratio between kinetic inductance and total inductance ($\frac{L_k}{L_k+L_l}$), (see section 3.1), $\Delta$ is NbN superconducting energy gap, $D(E_F)$ is density states of Fermi level, and $n_{qp}$ is quasiparticle density. Therefore, as expected from equation (7), the loss of the resonator will be mainly dependent on the temperature, and it will be inconsistent with sweeping power, which is what we see in our measurement at $T = 650$ mK and depicted in Fig. 4(a).

Table 2. Comparing the measured highest internal quality factor of each sample.

|  | $T$ (mK) | $f_r$ (GHz) | $Q_i$ (high power) | $Q_i$ (single photon) | $Q_c$ | $Q_L$ |
|---|---|---|---|---|---|---|
| **Sample 1** | 100 | 5.303 | $1.07 \times 10^6$ | $1.36 \times 10^5$ | 6378 | 6340 |
|  |  |  | $<n_{ph}> \sim 2.7 \times 10^4$ | $<n_{ph}> \sim 1$ |  |  |
| **Sample 2** | 26 | 5.952 | $9.76 \times 10^5$ | $9.57 \times 10^4$ | 2308 | 2252 |
|  |  |  | $<n_{ph}> \sim 2 \times 10^3$ | $<n_{ph}> \sim 1$ |  |  |

Table 2 compares the highest measured $Q_i$ of both samples at millikelvin temperatures during power sweep measurements. It shows that the $Q_i$ values are compatible with each other in single photon and high power regimes, even though each sample has been fabricated in a different timeline. These results confirm that thin NbN superconductor film and silicon substrate quality is approximately maintained regardless of the environmental situation and measurement setup.

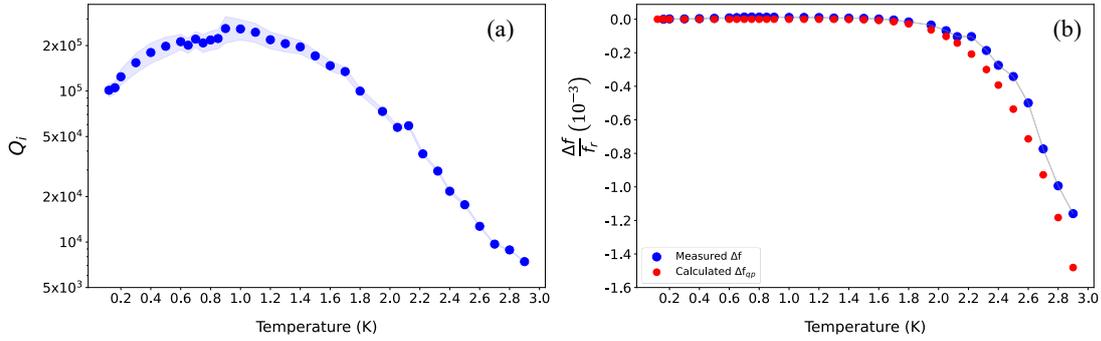

**Figure 5**. (a) $Q_i$ versus temperature at single photon regime. The shaded area corresponds to deviation/ error from the actual data $f_r = 5.952$ GHz. (b) Resonance frequency shift versus temperature. The blue and red circles show $\Delta f_{qp}$ and $\Delta f$ calculated quasiparticle loss frequency shift, and total loss frequency shift, respectively. These results are measured data of sample 2 in the ADR.

3.3 Temperature dependence of $Q_i$ and $f_r$
In the next step, we extracted the $Q_i$ and $f_r$ at single photon regime in the temperature ranges between $T = 100$ mK and 3K. The results show that at around $T \sim 1$ K, the $Q_i$ value is at its highest value, as shown in Fig. 5 (a). This behaviour is in good agreement with previous research on TiN and MoRe based microwave resonators [66, 67]. The same investigation has been done for the frequency shift $\Delta f$ of the resonator versus temperature; the data is shown in Fig. 5 (b). We observe that at $T \sim 1$ K, the resonance frequency experiences a noticeable red shift, the same temperature when we extracted the highest $Q_i = 2.57 \times 10^5 \pm 3.9 \times 10^4$. In more detail, when energy loss is mainly due to TLS defects, the resonance frequency shift also happens due to TLS loss (Fig. 5) which can be described by the below model [65]:

$$\Delta f_{TLS} = \frac{1}{Q_{TLS}^0} \frac{1}{\pi} \left[ R \left\{ \Psi \left( \frac{1}{2} + \frac{hf_r}{2\pi i k_B T} \right) \right\} - \ln \frac{hf_r}{2\pi k_B T} \right] \qquad (8)$$

where $\Psi$ is the digamma function. We can plot this function using data extracted from the previous fit in Fig. 5 (b). At higher temperatures, due to an increase in quasiparticle density, kinetic inductance increases, which leads to the below frequency red shift [65]:

$$\Delta f_{qp} = -\frac{1}{2} f_r \frac{\Delta L}{L} = -\frac{1}{2} \alpha f_r \frac{\Delta L_k}{L_k} = -\frac{1}{2} \alpha f_r \frac{\Delta}{k_B T} \frac{1}{\sinh \frac{\Delta}{k_B T}} \qquad (9)$$

In Fig.5 (b), we plot the measured $\Delta f$ relation with temperature and calculate frequency shifts obtained by equation (9) and equation (10). In this equation, we use the kinetic inductance ratio obtained from section 3.1. This theory agrees well with the measured data extracted from the resonator samples; the plot has been shown in Fig. 5 (b):

$$\Delta f \approx \Delta f_{TLS} + \Delta f_{qp} \qquad (10)$$

We see that $\Delta f_{qp} \sim \Delta f$ in higher temperatures; therefore, the main reason for resonance shifts in high temperatures is quasiparticle defects. Table 3 shows the highest and lowest $Q_i$ in a single photon regime for both samples measured during the temperature sweeps. It shows that $Q_i$ remains higher than $10^3$, although we measured our samples at high temperatures which are suspected to have high Cooper pair breaking rates.

Note that $Q_c$ and $Q_l$ values of sample 1 differ from the results in Table 2 because we did the wire bonding again due to the destruction of Aluminum wires during temperature sweep measurements. Wire bonding is mainly responsible for the impedance value of superconducting CPW. Therefore, as equation 1 shows, impedance mismatch affects the $Q_c$ and $Q_l$.

3.4 Evaluation of the CPWs under parallel magnetic fields

Finally, we examine the behaviour of the resonators when a static magnetic field is applied. The sample box (sample 1) is mounted in a cryogenic setup in the way that the NbN transmission line is positioned parallel to magnetic field $B_{//}$. We perform three measurements to characterise the magnetic field effect on the resonators. At first, we upsweep $B_{//}$ from 0 to 720 mT with 80 mT steps (see Fig. 6 (a)). The results of these measurements indicate that as the magnetic field is increased up to 720 mT, the resonance frequency experiences a redshift, and the dip value decreases because of Cooper pair breaking and emerging Abrikosov vortices.

Table 3: The highest and lowest internal quality factors of each sample in single photon regime measured during temperature sweep. Both of these measurements were performed in the ADR setup.

|  | Photon number | Max ($Q_i$) | Min ($Q_i$) | $Q_c$ | $Q_L$ |
|---|---|---|---|---|---|
| **Sample 1** | $<n_{ph}> \sim 1$ | $1.957 \times 10^5$ | $5.39 \times 10^4$ | 2418 | 2314 |
|  |  | $T = 700$ mK | $T = 2$ K |  |  |
| **Sample 2** | $<n_{ph}> \sim 1$ | $2.571 \times 10^5$ | $7.421 \times 10^3$ | 2142 | 2118 |
|  |  | $T = 1$ K | $T = 2.9$ K |  |  |

Trapped flux inside the NbN, as a type II superconductor, strongly depends on sample preparation and geometry. Moreover, existing and stability of these fluxes inside the superconductor, which affect resonator loss, come from two different forces, which appear due to the Meissner effect and Lorentz forces. First forces, due to the Meissner effect, a vortex is pushed toward the edges of the superconductor to expel from it. The second force happened when the vortex is attracted to the centre of the superconductor due to weak interactions of the Meissner screening current. The second one is proportional to the $B_{//}$, and at a high enough field, the vortex will exist and pin in a stable condition in a line of superconductor thin films [68, 69]. Based on theoretical investigation [68, 69], vortices in thin superconducting film for $B_{//} < B_a$ experience expulsion force, resulting in a more stable condition $B_a = \frac{\pi \phi_0}{4 t^2} = 161$ mT. Then vortices begin to align in a straight line when $B_a < B_{//} < B_{c1}$ and the edges of the thin film become vortices free $B_{c1} = 1.65 \times \frac{\phi_0}{t^2} = 341$ mT. For $B_{c1} < B_{//}$, the density of vortices increases in a way where vortices-vortices interaction should be taken into serious consideration where leading to a higher loss (Fig. 6 (a)). Here, $\phi_0$ and $t$ exhibit flux quantum and thickness, respectively. In the next step, we measure the effect of applying $B_{//}$ on the $f_r$ and $Q_i$ of CPW. As $B_{//}$ increases, $f_r$ experiences a redshift, as shown in Fig. 6 (b), due to the increased rate of Cooper pairs breaking. The frequency shift $\Delta f = f - f_r$ can be modelled as a parabolic decrease [70] as $\frac{\Delta f}{f_r} = -k \times B_{||}$, with $k = \frac{\pi}{48} \left( \frac{t^2 e^2 D}{\hbar k_B T_c} \right)$, which depends on resonator thickness

$t$, critical temperature $T_c$ and the electron diffusion constant $D$ in NbN. The electron diffusion constant is approximated to $D \sim 0.683$ cm$^2$ s$^{-1}$, consistent with the previous experiments [28, 71, 72].

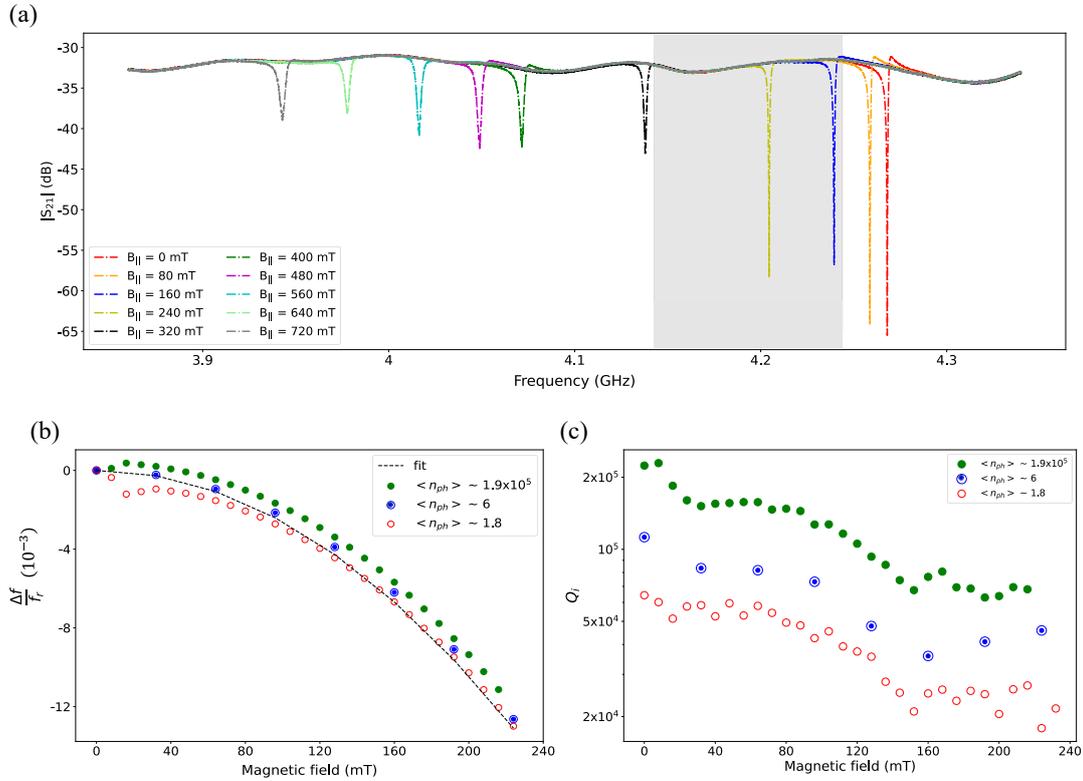

**Figure.6** (a) The $S_{21}$ versus frequency spectrum between $B_{\parallel} = 0$ T and 720 mT with 80 mT steps at $T = 100$ mK. The shaded area corresponds to frequency shift in magnetic field sweep in the range of $B_a < B_{//} < B_{c1}$. (b) Extracted frequency shift as a function of $B_{\parallel}$ (mT) with $<n_{ph}> = 1.9 \times 10^5$, 6 and 1.8. A fit of the measurement data with the equation $\frac{\Delta f_r}{f_{r0}} = -k \times B_{\parallel}$ which gives $k = 2.61 \times 10^{-1}$ T$^{-2}$ at $T = 100$ mK and $<n_{ph}> = 1.9 \times 10^5$, 6 and 1.8. (c) $Q_i$ as a function of $B_{\parallel}$ at $T = 100$ mK and $<n_{ph}> = 1.9 \times 10^5$, 6 and 1.8. The $Q_i$ remains above $10^4$ until $B_{//} = 240$ mT at all the mentioned photon numbers. These are measured data of sample 1 in the ADR cryogenic setup.

Figure 6 (c) demonstrates the effect of in-plane magnetic fields on the resonators $Q_i$ at $T = 100$ mK, and in $<n_{ph}> = 1.9 \times 10^5$, $<n_{ph}> = 6$ and $<n_{ph}> = 1.8$. It shows that by increasing the magnetic fields, $Q_i$ decreases. However, $Q_i$ remains above $10^4$ up to $B_{\parallel} = 240$ mT, which shows the resonator's effectiveness under applying magnetic fields. In superconducting resonators, the resonance frequency is determined by the RF surface impedance of the superconductor, which can be affected by flux creep and vortices pinning [68]. As magnetic vortices move and generate electrical resistance and heat, the surface impedance of the superconductor changes, which in turn can lead to a shift in the resonance frequency. The degree of resonance shift depends on several factors, including the strength of the applied magnetic field, the temperature, and the structure of the superconductor, including any defects and impurities or material disorder.

To further study the effect of in-plane magnetic fields on resonance frequency $f_r$, we reduced the steps in the magnetic field measurement to 8 mT. We performed an upsweep measurement in the range between 0 mT $< B_\parallel <$ 80 mT (Fig.7). We observed that $f_r$ experiences butterfly effects at low magnetic fields 0 mT $< B_\parallel <$ 32 mT, showing the mobility of the vortices which have a hysteresis effect in the conductivity of the NbN superconductors. Fig. 7(a) shows that the changes in the complex conductivity of planar superconducting circuits due to dissipative vortices motions lead to a jump in resonance frequency [73]. Indeed, this jump which occurs at magnetic fields ranging from 16 mT $< B_\parallel <$ 32 mT suggests a sudden movement or depinning of the vortices in the NbN superconductor. As the magnetic field increases, vortices tend to pin in the superconducting material. As shown in Fig. 7(b), we observed that, in a higher magnetic field range between 32 mT $< B_\parallel <$ 80 mT, the resonance frequencies decrease as the magnetic field increases implying that NbN thin film conductivity wasn't affected by the mobility of the vortices since no butterfly effect of the resonance frequency is detected [69,73].

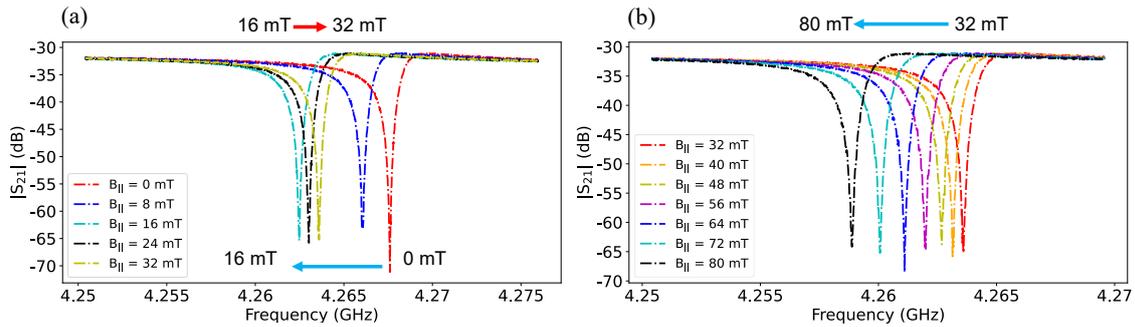

**Figure. 7** (a) Measured frequency spectrum for an upsweep measurement of between $B_\parallel$ = 0 mT to 32 mT with a step of 8 mT at temperatures $T$ = 100 mK. In this magnetic field range, we observe a resonance frequency jump at $B_\parallel$ = 16 mT. (b) Measured frequency spectrum for an upsweep measurement of the magnetic field between $B_\parallel$ = 32 mT to 80 mT, the resonance frequency shift as expected.

Moreover, to understand the effect of the solenoid on the $f_r$ of our sample, we increased the magnetic field's range to -160 mT $< B_\parallel <$ 160 mT. We expected that the $\Delta f$ plot should have a symmetrical shape around $B_\parallel$ = 0 mT where $\Delta f$ ($B_\parallel$ = 0) = 0. However, the maxima ($\Delta f$ = 0) shift towards negative fields $B_\parallel$ = -25.6 mT. This shift may happen due to the remaining magnetic field from the ADR magnet which is responsible for the adiabatic demagnetisation procedure. Besides, we observe a very high jump at $\Delta f$ ($B_\parallel$ = 32 mT) at $T$ = 100 mK (see Fig. 8). To identify the jump further, we performed temperature dependent measurements at $T$ = 100 mK, 500 mK, 1 K and 1.5 K to examine the effect of temperatures. As shown in Fig. 8, we observe a similar jump at all temperatures, which we presume is due to vortices trapped in the NbTi solenoid.

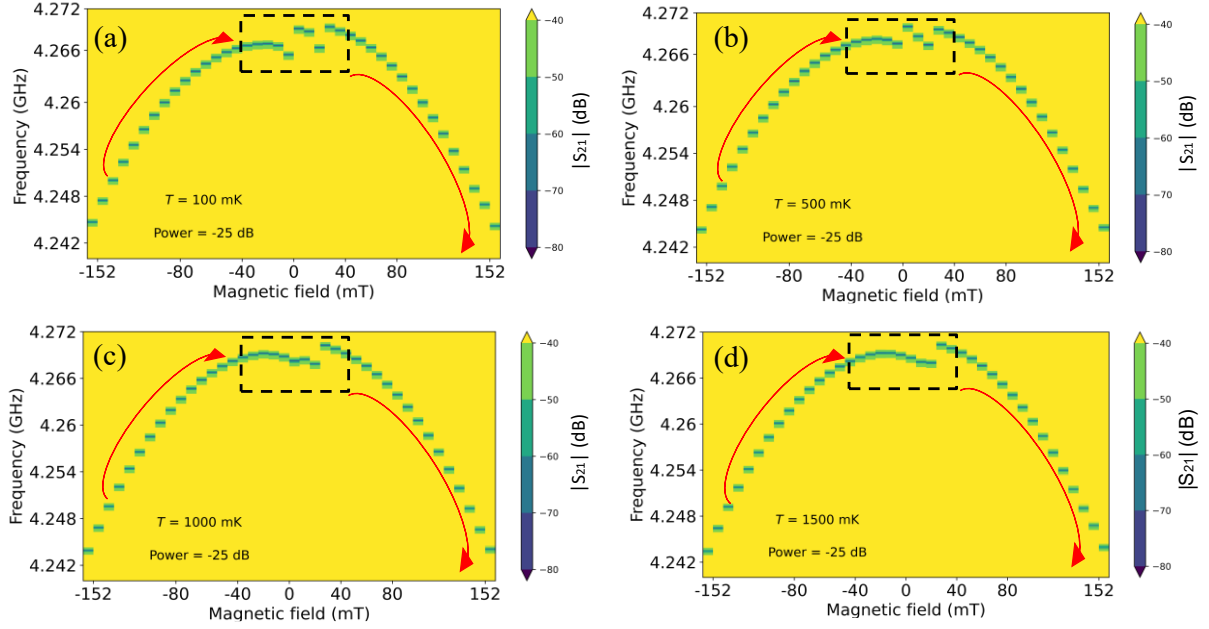

**Figure. 8** Measured frequency spectrum with magnetic field sweep raging between -160 < $B_∥$ (mT) < 160 with a field step of 8 mT at $T$ = (a) 100 mK, (b) 500 mK, (c) 1 K, (d) 1.5 K. Arrows indicate sweep direction.

To further study the progress on the fabrication procedure of NbN coplanar waveguide resonator, we compare our techniques with the most recent reports on NbN superconductor resonators in Table 4. We fabricated our CPW by NbN sputter deposition without heating, which causes a simpler and more robust fabrication procedure compared to atomic layer deposition (ALD) and heating sputtering. Note that the recent goal in quantum circuits research is to reduce the chip footprint, which we also move towards by design and fabrication of λ/4 resonators. To reduce the oxidation of NbN, we used $CF_4$ and Ar gases since they are less susceptible to introducing unwanted defects and material added to our superconducting thin film. However, more studies on how the etching gas influences the internal quality factors will be worth investigating.

Regarding design parameters, we choose 100 nm film thickness for two reasons: not only does 100 nm NbN superconducting thin film have a lower kinetic inductance, and is more robust against noises sources in the cryogenic measurement setup in comparison with thinner films, but it also gives a more robust internal quality factor after applying magnetic fields in comparison to bulk superconducting film [27,28]. Our resonator has stayed in the middle range of film thickness 4 nm < $t$ < 300 nm. However, finding a sweet spot in NbN film thickness will be worth more investigation. Moreover, we compare our resonator internal quality factor $Q_i$ in single photon and high power regimes to recent works to further understand the performance of NbN superconducting resonators.

Table 4. Comparison of NbN-based microwave coplanar waveguide resonator design, and fabrication.

| Article | Width | Gap | Thickness | Type | Metal deposition | Etching |
|---|---|---|---|---|---|---|
| **Carter et al [27]** | 12 µm | 8 µm | 300 nm | λ/4 | DC magnetron sputtering + 140 C⁰ heating | $SF_6/CHF_3$ |
| **Yu et al [28]** | 2 µm | 2 µm | 10 nm | λ/2 | DC magnetron sputtering | $O_2/SF_6$ |
| **Wei et al [63]** | -- | -- | 6 nm | λ/2 | DC magnetron sputtering | RIE |
| **Sheagren et al [29]** | > 2 µm | -- | 4 nm to 30 nm | -- | Atomic Layer deposition | $CF_4/CHF_3$ |
| **Foshat et al [this work]** | 4 µm | 2 µm | 100 nm | λ/4 | Sputtering | $CF_4/Ar$ |

The results are shown in Table 5. We find that at single photon regime, the $Q_i$ is in line with the latest research on superconducting NbN coplanar waveguide resonators, which confirms the reliability of the fabrication process and measurements at different magnetic fields and temperatures.

In terms of $Q_i$ at high power regime, our resonator has the second highest $Q_i$ value, even though we measured at far lower photon number and temperature than Carter, Faustin W., *et al* work. Our work in NbN based circuits is the first research demonstrating resonance frequency shift and internal quality factor behaviour as a function of temperature from the TLS loss to the quasi-particle loss ranges.

Table 5. Comparison of NbN-based coplanar waveguide resonator microwave response detail.

| Article | Temperature | Single photon $Q_i$ | High power $Q_i$ |
|---|---|---|---|
| **Carter et al [27]** | 200 mK | $4.2 \times 10^5$ | $1.7 \times 10^6$, $<n_{ph}> = 10^6$ |
| **Yu et al [28]** | 8 mK | $10^4$ | $10^6$, $<n_{ph}> = 10^5$ |
| **Wei et al [63]** | 30 mK | $7.5 \times 10^4$ | $5 \times 10^5$, $<n_{ph}> = 10^4$ |
| **Sheagren et al [29]** | 300 mK | -- | $> 10^5$ |
| **Foshat et al [this work]** | 100 mK | $1.35 \times 10^5$ | $1.07 \times 10^6$, $<n_{ph}> = 2.7 \times 10^4$ |

Meanwhile, although we obtained a high internal quality factor, we have not used artificial pinning in our circuit design, leading to a quicker and easier e-beam lithography procedure. These comparisons confirm that our approach in designing, fabricating, and characterising NbN superconducting CPW is a step towards developing high internal quality factor resonators useful for quantum technology from quantum computing to quantum sensing and metrology.

**Conclusion**

In summary, we reported the successful design, fabrication and microwave spectroscopy of arrays of CPWs based on superconducting NbN on silicon chips in the cryogenic environment. We studied resonator parameters as a function of microwave power, temperature, and magnetic fields. We studied TLS, quasiparticles, magnetic fields, and temperature-independent losses to identify the loss mechanism in our superconducting circuits. At low temperatures ($T = 100$ mK), the range in which TLS loss is the dominant reason for the rising loss in CPWs, the quality factor increases from $Q_i \sim 1.35 \times 10^5$ in the single photon regime to $Q_i \sim 1.07 \times 10^6$ at high powers. Sweeping temperature from $T = 100$ mK to 3 K in the single photon regime also confirms that TLS loss decreases when the TLS defects saturate in the energy bands. Moreover, in higher temperatures, the range in which quasiparticle loss is dominant, $Q_i$ reduces from $2.571 \times 10^5$ at $T \sim 1$ K to $7.421 \times 10^3$ at $T \sim 3$ K. To examine the suitability of such superconducting circuits for cQED at high magnetic fields, we studied the microwave response of our CPWs to in-plane magnetic fields $B_\parallel$. Our results showed that the peak of $S_{21}$ is stable in the magnetic field ranges between $B_\parallel = 0$ mT to 240 mT at $T = 100$ mK. We extracted $Q_i > 2 \times 10^4$ at single photon when $B_\parallel = 240$ mT. Our results shall help the development of superconducting microwave quantum circuits compatible with high magnetic fields necessary for applications in fault-tolerant quantum computing, quantum sensing, and quantum metrology.

**Acknowledgment:**
This work was supported in part by the Royal Society Research under Grant RGS/R2/222168, the Research Fellowship from the Royal Society of Edinburgh (K Delfanazari), the Innovate UK (QTools, grant No 79373 and FABU, grant No 50868), the EPSRC Quantum Computing Hub EP/T001062/1, and by the FET Open initiative from the European Union's Horizon 2020 program under N◦ 899561. The discussion with Dr A. Sultanov is greatly appreciated.

**Appendix:**

## S.1: Microwave spectroscopy setup

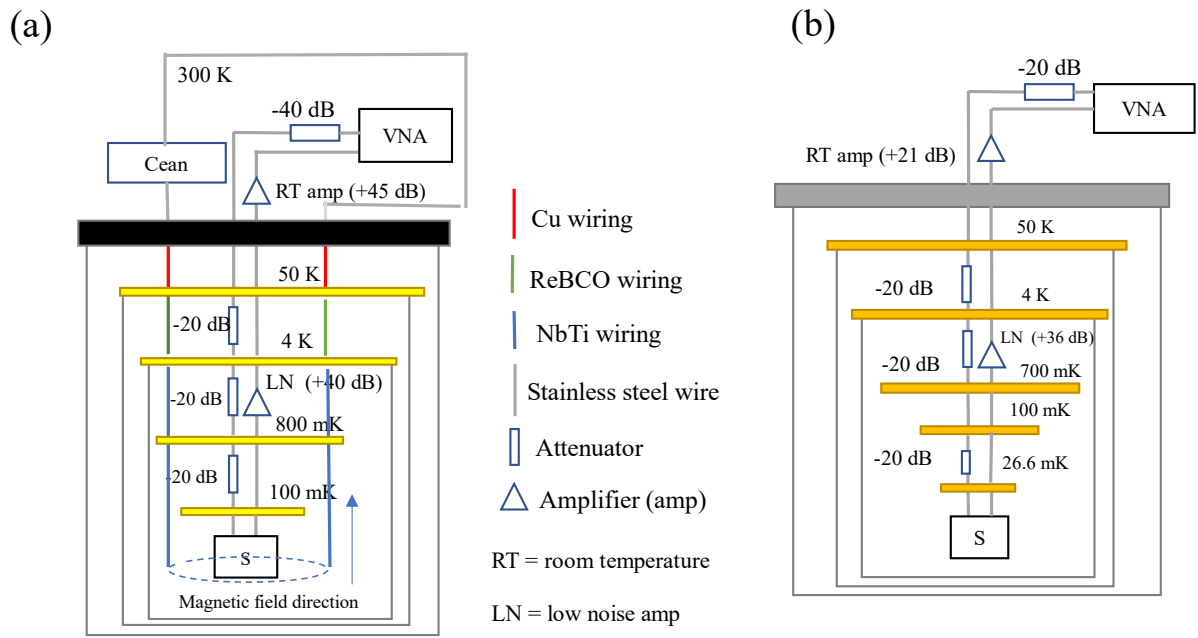

**FIG.S.1** (a) Adiabatic demagnetisation refrigerator (ADR) experimental setup schematic used to perform microwave and magnetic field measurements with NbTiN superconducting solenoid, Cean is the current source of NbTi solenoid. (b) Schematic of the dilution refrigerator used to perform microwave measurements.